\documentclass{aa}  

\usepackage{graphicx}
\usepackage{xcolor}
\usepackage{txfonts}
\usepackage{amsmath}    
\usepackage{amssymb}    
\usepackage{ulem}       
\usepackage{natbib,twoopt}
\bibpunct{(}{)}{;}{a}{}{,}             
\usepackage[breaklinks=true]{hyperref}
\hypersetup{
    colorlinks=true,
    linkcolor=blue,
    filecolor=magenta,      
    urlcolor=cyan,
    citecolor=blue
   }
\newcommand\blfootnote[1]{%
  \begingroup
  \renewcommand\thefootnote{}\footnote{#1}%
  \addtocounter{footnote}{-1}%
  \endgroup
}
\begin{document}

   \title{Complementary cosmological simulations}


    \author{G\'abor R\'acz\inst{1,2},
            Alina Kiessling\inst{1},
            Istv\'an Csabai\inst{2},
            \and
            Istv\'an Szapudi\inst{3}
          }
    \institute{Jet Propulsion Laboratory, California Institute of Technology, 4800 Oak Grove Drive, Pasadena, CA, 91109, USA\\
                \email{gabor.racz@jpl.nasa.gov}
        \and
                Department of Physics of Complex Systems, ELTE E\"{o}tv\"{o}s Lor\'and University, 
                Pf. 32, H-1518 Budapest, Hungary\\
         \and
             Institute for Astronomy, University of Hawaii, 2680 Woodlawn Drive, Honolulu, HI, 96822, USA\\
             }

    \date{Received October 13, 2022; accepted January 30, 2023}

 
  \abstract
   {Cosmic variance limits the accuracy of cosmological $N$-body simulations, introducing bias in statistics such as the power spectrum, halo mass function, or the cosmic shear.}
   {We provide new methods to measure and reduce the effect of cosmic variance in existing and new simulations.}
   {We ran pairs of simulations using phase-shifted initial conditions with matching amplitudes. We set the initial amplitudes of the Fourier modes to ensure that the average power spectrum of the pair is equal to the cosmic mean power spectrum from linear theory.}
   {The average power spectrum of a pair of such simulations
   remains consistent with the estimated nonlinear spectra of the state-of-the-art methods even at late times. We also show that the effect of cosmic variance on any analysis involving a cosmological simulation can be estimated using the complementary pair of the original simulation. To demonstrate the effectiveness of our novel technique, we simulated a complementary pair of the original Millennium run and quantified the degree to which cosmic variance affected its the power spectrum. The average power spectrum of the original and complementary Millennium simulation was able to directly resolve the baryon acoustic oscillation features.}
   {}

   \keywords{large-scale structure of Universe --
            dark matter --
            methods: numerical
               }

   \maketitle
%

\section{Introduction}
\blfootnote{\copyright 2022. All rights reserved.}
Large-scale galaxy surveys are important probes to test the standard Lambda Cold Dark Matter ($\Lambda$CDM) model of cosmology in addition to alternative cosmologies. Past surveys, such as the Automated Plate Measurement (APM) galaxy survey \citep{1990MNRAS.243..692M, 1994MNRAS.267..323B}, 2dF \citep{2001MNRAS.327.1297P}, Two Micron All-Sky Survey (2MASS) \citep{2001astro.ph..9403A}, Sloan Digital Sky Survey (SDSS) \citep{2004ApJ...606..702T}, Dark Energy Survey (DES) \citep{2005astro.ph.10346T}, Dark Energy Spectroscopic Instrument (DESI) \citep{2016arXiv161100036D}, and the Panoramic Survey Telescope and Rapid Response System (Pan-STARRS) \citep{2016arXiv161205560C}, have provided valuable data regarding the evolution of the matter in our Universe. The upcoming Rubin \citep{2009arXiv0912.0201L}, Roman \citep{2012arXiv1208.4012G}, Spectro-Photometer for the History of the Universe, Epoch of Reionization, and ices Explorer (SPHEREx) \citep{2014arXiv1412.4872D} and Euclid \citep{2020A&A...643A..70T} surveys will greatly extend these observations by mapping unprecedented volumes with previously unseen etendue.
Constraints on cosmological parameters are derived by comparing observations and theoretical predictions using statistical quantities such as the power spectrum, halo mass function, or the cosmic-shear two-point correlation function. Cosmological $N$-body simulations are widely used for calculating these quantities at late times \citep[e.g.,][]{1991MNRAS.251..575V,2006MNRAS.371.1205R,2001MNRAS.321..372J,2012A&A...542A.126C}. To answer the demands of the large surveys, simulations were run to calculate the nonlinear evolution of the 3D matter distribution inside large volumes with high precision over the years, such as the Millennium run \citep{2005Natur.435..629S}, Millennium-II \citep{2009MNRAS.398.1150B}, Bolshoi \citep{2011ApJ...740..102K}, Millennium-XXL \citep{2012MNRAS.426.2046A}, MultiDark \citep{2016MNRAS.457.4340K}, the Euclid flagship simulation \citep{2017ComAC...4....2P}, and the Outer Rim Simulation \citep{2019ApJS..245...16H}. The results of these simulations have been used in hundreds of research  projects and are still regularly used today.\\

Both observations and cosmological simulations are affected by cosmic variance, because they are both only sampling a finite volume of the cosmic density field \citep{2010MNRAS.407.2131D, 2011ApJ...731..113M, 2016JCAP...04..047S}. The most straightforward method to estimate the effects of cosmic variance is to use an ensemble of cosmological simulations such as the LasDamas \citep{2009AAS...21342506M}, the Indra \citep{2021MNRAS.506.2659F} or the ABACUSSUMMIT \citep{2021MNRAS.508.4017M} simulation suites. By comparing the results of hundreds of simulations, the cosmic mean and variance can be directly calculated for a given quantity. Paired \citep{2016PhRvD..93j3519P} and paired-and-fixed \citep{2016MNRAS.462L...1A} initial conditions can effectively reduce the cosmic variance in simulations. Detailed descriptions of these initial conditions are presented in Sect.~\ref{sec:InvertedICs} and Sect.~\ref{sec:FixedAndPairedICs}, respectively. By running just two simulations, the results of the paired-and-fixed initial conditions method can closely match the average of the ensembles of 300 simulations for density distribution, power spectrum, and bispectrum. Although the statistics of the average dark matter clustering are precise in these simulations, the covariances from paired-and-fixed simulations are suppressed, and as a consequence, these cannot be used for generating mock galaxy catalogs as shown in \cite{2020MNRAS.496.3862K}. \cite{2019A&A...631A.160H} pointed out that reducing the effects of cosmic variance on the cosmic shear covariance calculation is possible with matched pairs of simulations.\\

Our aim is to extend the techniques shown above by proposing alternative ways to generate initial conditions for cosmological simulations. The outline of this paper is as follows:
In Sect.~\ref{sec:ICs}, we overview the traditional
algorithm used to generate initial conditions and propose modifications to this in order to reduce the cosmic variance.
Then, in Sect.~\ref{sec:Simulations}, we compare the different simulations that started from the modified initial conditions.
In Sect.~\ref{sec:Millennium}, we demonstrate our new best performing method on the original Millennium simulation by running its complementary pair. Using this new simulation, we estimate the effect of cosmic variance on the original Millennium power spectrum.
Finally, we summarize our results.\\

\section{Initial conditions for cosmological simulations}
\label{sec:ICs}
The main goal of initial-condition creation is to represent a $\delta(\vec{x}) = \rho(\vec{x})/\overline{\rho} -1$ overdensity field with discrete particles. For cosmological simulations, this field should be consistent with the $P_{target}(k)$ initial power spectrum. The first step in this process is to generate the Fourier transform of the overdensity field \citep{1988csup.book.....H, 2005ApJ...634..728S}:
\begin{equation}
    \delta(\vec{k}) =  \int \delta(\vec{x}) e^{-i\vec{k}\vec{x}} d^3x.
\end{equation}
As this is a complex field, it can be written as
\begin{equation}
    \delta(\vec{k}) = a(\vec{k}) + i \cdot b(\vec{k}) = A(\vec{k})\cdot e^{i\varphi(\vec{k})},
\end{equation}
and it should satisfy the usual Hermitian constraints, because its inverse Fourier transform is a real field. For a random Gaussian realization of a target $P_{target}(k)$ initial power spectrum, $a(\vec{k})$ and $b(\vec{k})$ should be drawn independently from a Gaussian distribution \citep{1997astro.ph.12217K}
\begin{equation}
    P_G(a)da = \frac{1}{\sigma(k)\sqrt{2\pi}}e^{- \frac{a^2}{2\sigma(k)^2}}da,
\end{equation}
for every resolved $\vec{k}$ wavenumber vector. Equivalent to this is to generate the magnitude $A(\vec{k})$ from the Rayleigh distribution
\begin{equation}
        P_R(A)dA=\frac{A}{\sigma(k)^2}e^{-\frac{A^2}{2\sigma(k)^2}}dA,
\end{equation}
and the corresponding phase $\varphi(\vec{k})$ is chosen randomly from a uniform distribution on a range between 0 and $2\pi$ independently \citep{2005ApJ...634..728S}, where $\sigma(k)^2=VP_{target}(k)/(2\pi)^3$, $V$ is the volume of the simulation, and $k=|\vec{k}|$.  After the initial field is generated in Fourier space, the $\delta(\vec{x})$ overdensity field is calculated using the inverse Fourier transform.
In the last step of the initial-condition generation, this Eulerian density is transformed into a Lagrangian representation of particles using the Zel'dovich approximation \citep{1970A&A.....5...84Z} or the second-order Lagrangian perturbation theory (2LPT) \citep{2006MNRAS.373..369C, 2010MNRAS.403.1859J}. The power spectrum of the traditional initial conditions,
\begin{equation}
    P_{IC}(k_1) = \frac{(2\pi)^3}{V}\delta_{\vec{k}_1,\vec{k}_2}\left<\delta(\vec{k}_1)\delta^*(\vec{k}_2)\right> = \frac{(2\pi)^3}{V}\left<A(\vec{k}')^2\right>_{|\vec{k}'|=k_1}
,\end{equation}
will not be equal to $P_{target}(k)$, because the initial amplitudes are drawn from a Rayleigh distribution, and this is the main source of the cosmic variance in simulations. In this article, we use $\left<\right>_{|\vec{k}'|=k}$ to denote the average value of the $|\vec{k}'|=k$ surface in Fourier space. In the rest of this paper, we discuss modifications to this traditional method and show how these modifications can reduce the effects of cosmic variance in cosmological $N$-body simulations.

\subsection{Inverted initial conditions}
\label{sec:InvertedICs}
This method was proposed by \cite{2016PhRvD..93j3519P} and can be used to create a density-inverted counterpart to an existing initial condition. This can be achieved by running the same initial condition generator again with the same parameters and random seed, but with $\varphi(\vec{k})$ phases shifted by $\pi$. As a consequence of this phase shift, the initial $\delta(\vec{x})$ overdensity fields are inverted in the counterpart, that is, underdensities are substituted for overdensities and vice versa.
The simulations that start from the pairs of these initial conditions are called Paired simulations. The power spectrum of the original and the inverted initial condition are the same, because they only differ in the phases. Despite the fact that the pair have the same initial power spectrum, the cosmic variance can still be reduced by averaging the results of paired simulations, as the averaging cancels phase correlations that emerge from the late-time nonlinear evolution of the density field \citep{2018ApJ...867..137V}.

\subsection{Fixed and paired-and-fixed initial conditions}
\label{sec:FixedAndPairedICs}
The next step in reducing cosmic variance was developed by \cite{2016MNRAS.462L...1A}. In their approach, the amplitudes of the initial conditions are fixed to the expected mean value by setting
\begin{equation}
    A(\vec{k}) = \sqrt{\frac{VP_{target}(|\vec{k}|)}{(2\pi)^3}}.
    \label{eq:FixedA}
\end{equation}
The simulations that start from this initial condition are called Fixed simulations. A further reduction of the cosmic variance can be
achieved by combining this technique with the paired simulation method described in Sect. \ref{sec:InvertedICs}  \citep{2016MNRAS.462L...1A}. These paired-and-fixed (PF) simulations can reduce the variance of the power spectrum by factors as large as $10^6$. A detailed description of the statistical properties of the paired and PF simulations can be found in \cite{2018ApJ...867..137V}. Although the PF method is very effective in reducing the variance of the power spectrum, the halo mass function, and the bispectrum, \cite{2020MNRAS.496.3862K} found that the covariances are significantly suppressed, in contrast to the series of traditional simulations with normal Gaussian initial conditions. The main source of this bias is the fact that the initial conditions are missing the random amplitude fluctuations. As this is a well known and very effective method for reducing the cosmic variance, we use the PF simulations as a reference when we compare the different power spectra in this paper.

\subsection{Matched pairs}

An independent approach was used by \cite{2019A&A...631A.160H} to reduce cosmic variance in cosmic-shear covariance calculations. These authors generated a large ensemble of independent initial conditions and chose two from the set that had an average that was closest to $P_{target}(k)$. This pair of initial conditions was used as the starting point for a pair of simulations. Unlike the Paired simulations described in Sect.~\ref{sec:FixedAndPairedICs}, the effects of phase correlations are not canceled out during the averaging of these simulations, as the $\varphi(\vec{k})$ phases are completely independent in these pairs.

\subsection{Paired-and-mean initial conditions}

The first new method we propose in this article is the paired-and-mean (PM) initial conditions. According to this method, the initial $\delta(\vec{k})$ field is generated in the traditional way in the first step, but the initial amplitudes are scaled by
\begin{equation}
    A_{PM}(\vec{k}) = A(\vec{k})\cdot\sqrt{\frac{VP_{target}(k)}{(2\pi)^3\left<A(\vec{k}')^2\right>_{|\vec{k}'|=k}}},
\end{equation}
where $A_{PM}(\vec{k})$ are the new PM amplitudes used in the initial condition, and $A(\vec{k})$ are the amplitudes generated with the traditional technique. This transformation guarantees that the initial power spectrum is equal to $P_{target}(k)$, while there are still fluctuations in the amplitudes. We propose that these simulations be run in inverted pairs in order to reduce the effect of phase correlations.

In practice, PM initial conditions can be generated using traditional initial-condition generators such as NGen\_IC \citep{2015ascl.soft02003S} or 2LPTic \citep{2012ascl.soft01005C, 2006MNRAS.373..369C} using the following target power spectrum:
\begin{equation}
    P_{input, PM}(k) = \frac{P_{target}(k)^2}{P_{IC}(k)},
\end{equation}
where $P_{IC}(k)$ is the power spectrum of a traditional initial condition generated with the same code, parameters, and random seed. As the Zel'dovich or 2LPT approximations also have an effect on the final power spectrum \citep{2006MNRAS.373..369C}, $P_{IC}(k)$ should be calculated from the generated particle distribution.

\subsection{Complementary initial conditions}

Our aim is to use pairs of simulations as part of this new technique, where the initial condition of the first simulation is generated in the traditional way, while the second has modified initial amplitudes and phases. We refer to the former as the "original" simulation, and the latter as the "complementary" simulation. We choose the initial $A_{C}(\vec{k})$ values in the complementary initial condition so that the average of the two initial $P(k)$ matches the $P_{target}(k)$ power spectrum. This constraint can be written as
\begin{equation}
    \left<A_C(\vec{k}')^2\right>_{|\vec{k}'|=k} = \frac{2V}{(2\pi)^3}P_{target}(k) - \left<A(\vec{k}')^2\right>_{|\vec{k}'|=k},
    \label{eq:complementarycriteriaI}
\end{equation}
where $A_C(\vec{k})$ is the amplitude in the complementary initial condition at $\vec{k}$ wavenumber. This can only be satisfied for the $k$ wavenumbers, where 
\begin{equation}
    2 \cdot P_{IC}(k) < P_{target}(k),
    \label{eq:complementarycriteriaII}
\end{equation}
because $\left<A_C(\vec{k}')^2\right>_{|\vec{k}'|=k}$ must be a positive. Although this cannot be done for the entire $|\vec{k}| = k$ surface, this technique can lead to a significant reduction in the variance of the power spectrum by compensating amplitudes that satisfy this criterion. There are an infinite number of $A_C(\vec{k})$ fields that satisfy Eq.~\ref{eq:complementarycriteriaI} when Eq.~\ref{eq:complementarycriteriaII} is true. In this study, we chose to generate these fields with the following formula:
\begin{equation}
\begin{split}
    &A_{C}(\vec{k}) \biggr\rvert_{2 \cdot P_{IC}(k) < P_{target}(k)} = \\
    &= A(\vec{k})\sqrt{\frac{VP_{target}(k)}{(2\pi)^3\left<A(\vec{k}')^2\right>_{|\vec{k}'|=k}} \left(2 - \frac{(2\pi)^3\left<A(\vec{k}')^2\right>_{|\vec{k}'|=k}}{VP_{target}(k)}\right)} = \\
    &= A_{PM}(\vec{k})\cdot\sqrt{\left(2 - \frac{(2\pi)^3\left<A(\vec{k}')^2\right>_{|\vec{k}'|=k}}{VP_{target}(k)}\right)}
    \end{split}.
\end{equation}
 For the modes that cannot be compensated, instead of setting them at zero, we chose
\begin{equation}
    \begin{split}
    A_{C}(\vec{k})\biggr\rvert_{2 \cdot P_{IC}(k) \geq P_{target}(k)} &= A(\vec{k})\frac{VP_{target}(k)}{(2\pi)^3\left<A(\vec{k}')^2\right>_{|\vec{k}'|=k}}  = \\
    &= A_{PM}(\vec{k})\cdot\sqrt{\frac{VP_{target}(k)}{(2\pi)^3\left<A(\vec{k}')^2\right>_{|\vec{k}'|=k}}}
    \end{split}
\end{equation}
in order to avoid introducing any strong unwanted  beat-coupling effects \citep{2006MNRAS.371.1188H}. The advantage of this choice is that the complementary initial condition can be generated with the usual codes simply by setting the target power spectrum to
\begin{equation}
\begin{split}
    P_{input, C}(k)\biggr\rvert_{2 \cdot P_{IC}(k) < P_{target}(k)} &= 2\frac{P_{target}(k)^2}{P_{IC}(k)}-P_{target}(k) = \\
    &= P_{input, PM}(k) \cdot \left(2 - \frac{P_{IC}(k)}{P_{target}(k)}\right)
\end{split}
,\end{equation}
\begin{equation}
    P_{input, C}(k)\biggr\rvert_{2 \cdot P_{IC}(k) \geq P_{target}(k)} = P_{input, PM}(k) \cdot \left(\frac{P_{target}(k)}{P_{IC}(k)}\right),
\end{equation}
with the same random seed and parameters, but shifted phases.
With this new method, it is possible to provide complementary simulations for existing simulations. By averaging the results of these pairs, a significant reduction in the cosmic variance can be expected, which is similar to the results of the PF simulations. We call these pairs original-complementary (O-C) pairs. It is possible to generate initial conditions without a phase shift, and these are called original-complementary amplitude (O-CA) pairs.\\

\section{Comparing the different simulation methods}
\label{sec:Simulations}

\begin{figure}
    \centering
    \includegraphics[width=1.05\hsize]{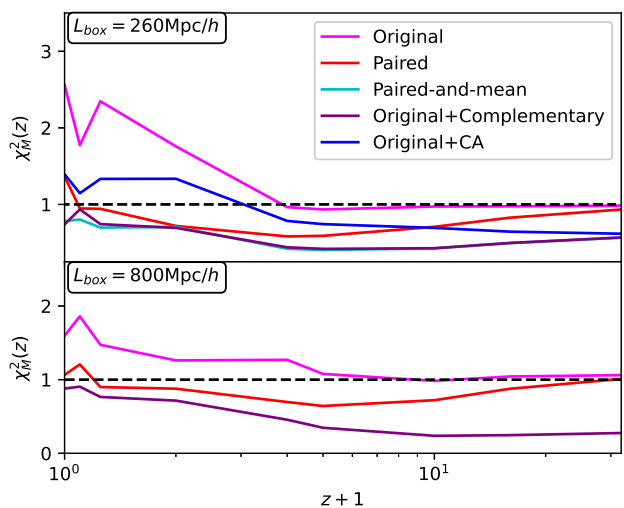}
    \caption{Reduced chi-squared statistics of the power spectrum. If this quantity is close to one, the power spectrum of the given method is consistent with the PF method. While the original simulation becomes inconsistent with the cosmic average power spectrum at late times, the average power spectrum of the PM and O-C simulations is a good match to the PF results. We used all available $k\leq1.5h/\textnormal{Mpc}$ modes when we calculated these $\chi^2_{\nu}(z)$ functions. }
    \label{fig:ReducedChi2}
\end{figure}

\begin{table*}[h!]
\caption{Parameters of simulations}             
\label{tab:simulations}      
\centering                          
\begin{tabular}{c | c c c c }        
\hline\hline                 
Simulation sets & L260\_N20M & L800\_N531M & MillenniumG4 & NewMillennium \\    
\hline                        
IC generator & 2LPTic & N-GenIC & L-GenIC & L-GenIC \\      
Initial redshift & 31 & 127 & 127 & 127 \\
Final redshift & 0 & 0 & 0 & 1 \\
$L_{box}[\textnormal{Mpc}/h]$ & $260$ & $800$ & $500$ & $500$  \\
$N_{part}$ & $270^3$ & $810^3$ & $2160^3$ & $2160^3$ \\
Cosmology & Planck2018 & Planck2018 & Millennium  & Millennium \\
Initial Conditions & O, P, PF, PM, CA, C & O, P, PF, C & O, C & O, C \\
\hline                                   
\end{tabular}
\tablefoot{The abbreviations of the different initial condition types are: O -- original; P -- phase shifted by $\pi$; PF -- paired-and-fixed; PM -- paired-and-mean (PM); CA -- complementary amplitudes without phase shift; C -- complementary.}
\end{table*}

\begin{figure}
    \centering
    \includegraphics[width=1.05\hsize]{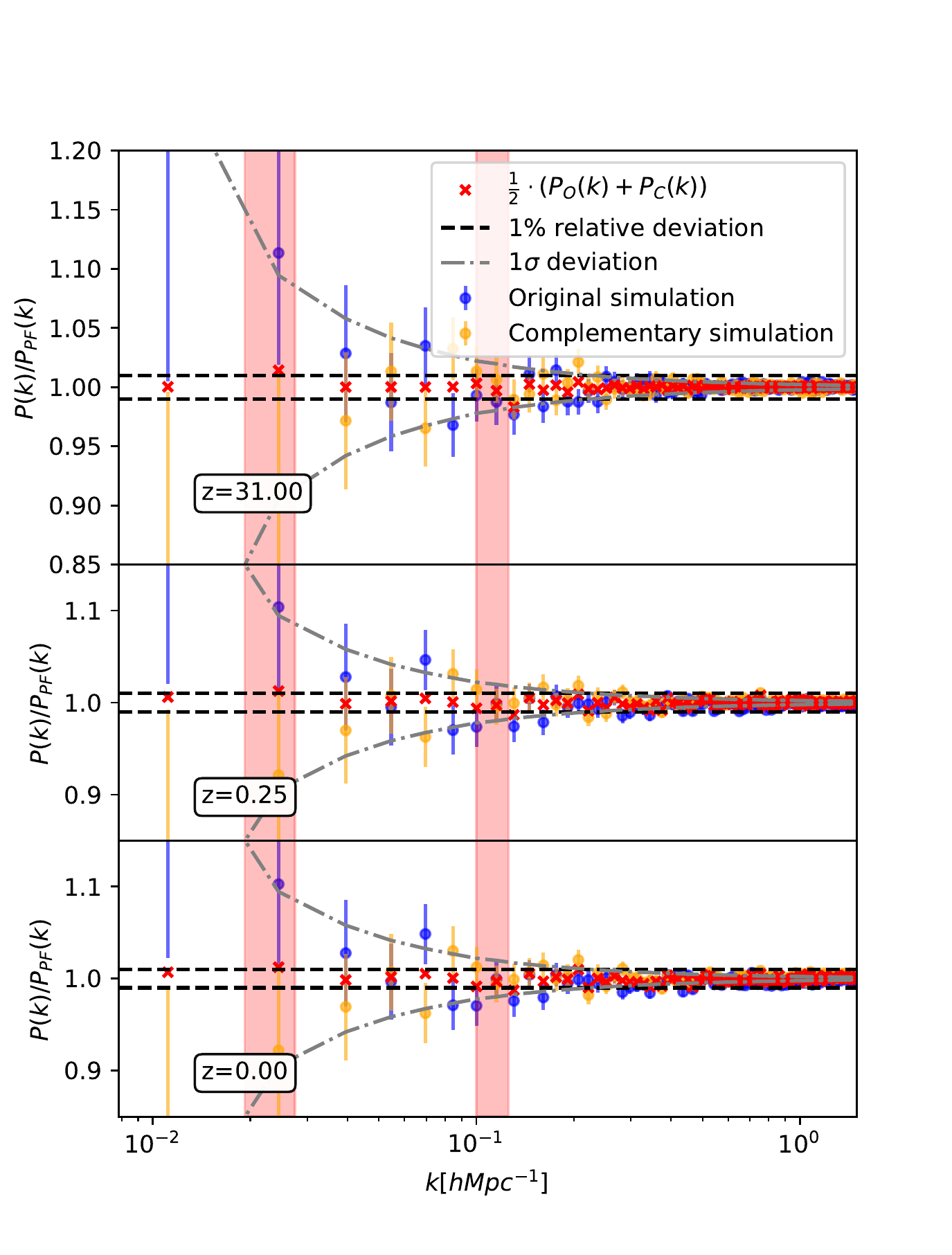}
    \caption{Ratio of the O-C and PF power spectra in the $L=800\textnormal{Mpc}/h$ simulation set. The red shaded regions show the ranges where uncompensated modes were present in the complementary simulation. 
    Although these modes were not fully compensated in the average, we included these in the $\chi^2_{\nu}(z)$ statistics plotted in Fig.~\ref{fig:ReducedChi2}. }
    \label{fig:OnePercentPk}
\end{figure}

To test the effects of the new initial-condition-generation techniques, we ran four sets of cosmological $N$-body simulations. A summary of the simulations is provided in Table~\ref{tab:simulations}. All simulations were run with GADGET-4 \citep{2021MNRAS.506.2871S}, and two had the Planck 2018 cosmological parameters \citep{2020A&A...641A...6P}, while two used the parameters of the Millennium run \citep{2005Natur.435..629S}. In the $L_{box} = 260.0\textnormal{Mpc}/h$ and $L_{box} = 800.0\textnormal{Mpc}/h$ sets of simulations, there were a few modes where Eq.~\ref{eq:complementarycriteriaII} was not satisfied: only one mode in the $L_{box} = 260.0\textnormal{Mpc}/h$ set and two modes in the $L_{box} = 800.0\textnormal{Mpc}/h$ series. To compare the power spectra of the different simulation techniques, we used the PF power spectrum as a reference in the $L_{box} = 260.0\textnormal{Mpc}/h$  and $L_{box} = 800.0\textnormal{Mpc}/h$ simulation sets, and calculated the
\begin{equation}
    \chi^2_{\nu}(z) = \frac{1}{\nu}\sum\limits_i \frac{(P_{M}(k_i,z)-P_{PF}(k_i,z))^2}{\sigma^2(k_i,z)}
    \label{eq:ReducedChi2}
\end{equation}
reduced $\chi^2$ for each $M$ method, where $\nu$ is the total number of $k_i$ bins, $\Delta N_{m,i}$ is the number modes per bin, and
\begin{equation}
    \sigma(k_i,z) = \sqrt{\frac{2}{\Delta N_{m,i}}}P(k)
\end{equation}
is the expected statistical error originating from the sample variance \citep{2016JCAP...04..047S}. The choice of binning has a small impact on the calculated $\chi^2_{\nu}(z)$ quantity, as the contribution of each bin in Eq.~\ref{eq:ReducedChi2} is weighted by $\Delta N_{m,i}$. The calculated reduced chi-squared statistic as a function of the redshift is plotted in the top panel of Fig.~\ref{fig:ReducedChi2}. The same quantity was calculated for the $L=800\textnormal{Mpc}/h$ simulation set, and this can be seen in the bottom panel of the same figure. Although the original simulations in both sets were close to $\chi^2_O\simeq1$ at early times, this value increased quickly as the matter field evolved. The power spectra from the P and O-CA simulations perform significantly better than the original simulations, but their $\chi^2$ values are still above 1 at a redshift of $z=0$. As expected, the PM and O-C simulations perform best during these tests: the $\chi_{PM}^2$ and $\chi_{O-C}^2$ values remain below 1 throughout the entire simulated redshift range. An additional visualization of the effect of the new method on the power spectrum is shown in Fig.~\ref{fig:OnePercentPk}, which shows the ratio of the O-C pair and the PF power spectrum for the $L=800\textnormal{Mpc}/h$ simulations with $1\%$ and $1\sigma$ deviations. The complementary simulation method significantly reduces the cosmic variance of the power spectrum at all simulated redshifts, and has a sub-percent accuracy for all compensated modes, even at the largest resolved scales.

The final simulation set contains a single O-C pair. Each simulation in the pair has $L=500\textnormal{Mpc}/h$ linear size with  10 billion particles and requires 0.74 million CPU hours to run. We used this set in Sect.~\ref{sec:Millennium} to measure the differences between two independent O-C realizations of the same cosmology. We call this simulation pair "NewMillennium".

\begin{figure*}
        \includegraphics[width=0.95\hsize]{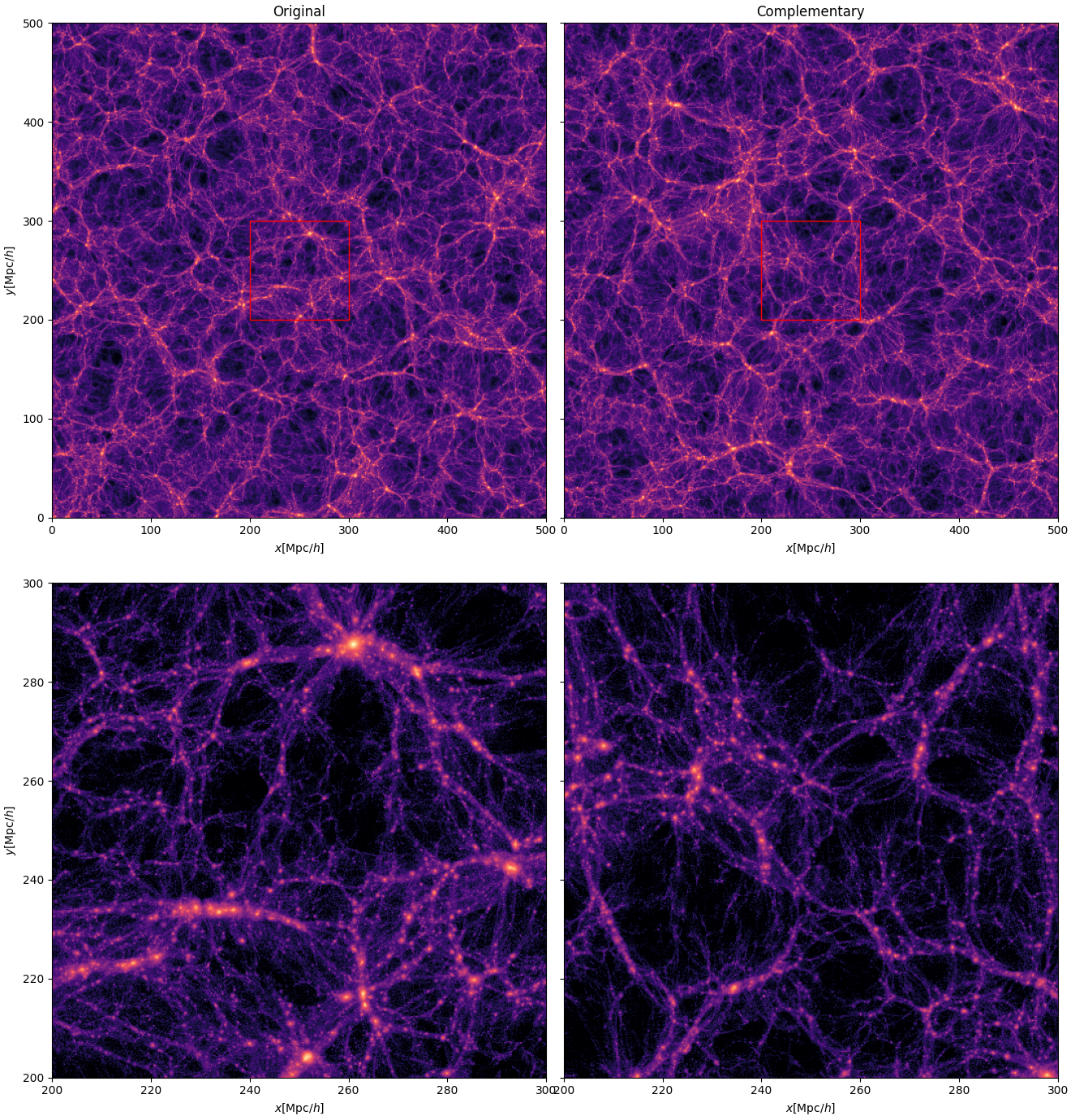}
      \caption{Dark matter density field of the Millennium run (\textbf{left}) and its complementary pair (\textbf{right}) at $z=0$ redshift. The bottom row is a zoom onto the region outlined by the red boxes in the figures in the top row. As the $\varphi(\vec{k})$ phases are shifted by $\pi$ in the initial condition, a region that collapses to a halo in the original Millennium run tends to expand into a void in the corresponding complementary pair and vice versa, which is similar to the paired simulation method.}
      \label{Fig:DensityField}
 \end{figure*}

\begin{figure*}
        \includegraphics[width=0.95\hsize]{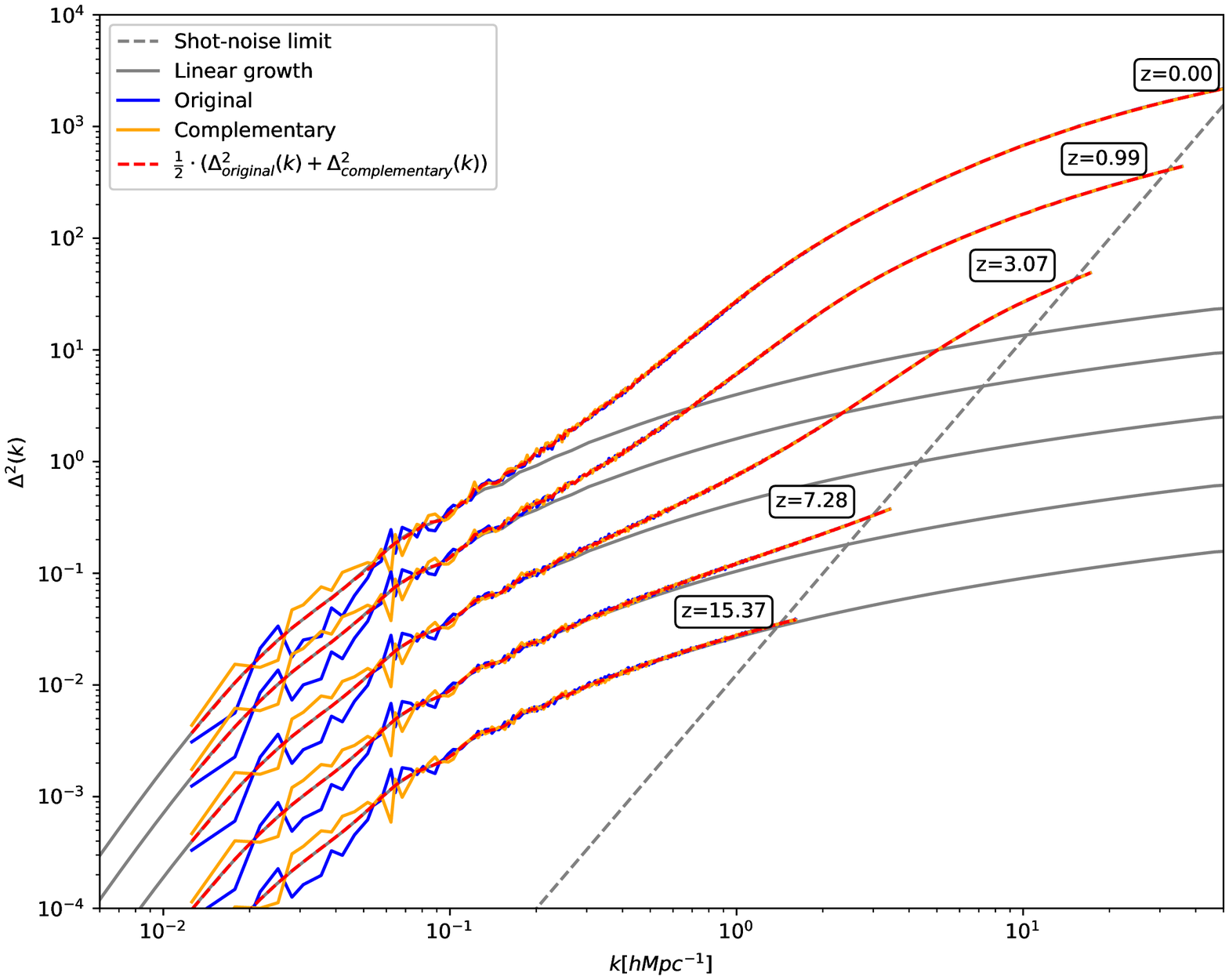}
      \caption{The $\Delta^2(k) = P(k)k^3/(2\pi^2)$ dimensionless power spectrum of the Millennium run and its complementary pair at $z=0$ redshift. The dashed line represents the shot-noise limit. We plot the linear power spectrum with a gray solid line for the different redshifts. The average dimensionless power spectrum of the pair initially matched this linear spectrum. The complementary pair compensates for the fluctuations around the linear input spectrum due to Rayleigh sampling  at all scales initially.}
      \label{Fig:DimlessPk}
\end{figure*}

\section{Complementary Millennium-run}
\label{sec:Millennium}

To further demonstrate the effectiveness of the O-C pair method in reducing cosmic variance, we planned to run the complementary pair of the original Millennium run. This dark matter only TreePM simulation contains $2160^3$ particles in a periodic box with a linear size of $L_{box} = 500Mpc/h$, and follows the evolution of cosmic structures from redshift $z=127$ to $z=0$ with more than 10 000 time steps. The cosmological parameters are the following: $\Omega_m = 0.25$, $\Omega_b = 0.045$, $h = 0.73$, $\Omega_\Lambda = 0.75$, $n = 1$, and $\sigma_8 = 0.9$. As using the same Lean-Gadget-2 code \citep{2005Natur.435..629S, 2005MNRAS.364.1105S} that was used to run the original simulation is not possible on modern systems, we chose to use the more recent Gadget-4 for this task. To estimate the effects of using a different simulation code, we re-simulated the small volume version of the Millennium run called "Milli-Millennium" with Gadget-4. We find a percent-level discrepancy in the power spectrum at the final state between the results of the LeanGadget-2 and Gadget-4 simulations.
We conclude that these differences emerge from a restart of the original simulation, and from some unknown differences in the default setups for the transition between the particle mesh and tree force calculations. As we were unable to minimize the effects emerging from using different versions of the same code, we re-simulated the original Millennium run with Gadget-4. We used the resources of the Texas Advanced Computing Center (TACC) to run the Millennium simulation and its complementary pair.

\begin{figure}
    \centering
    \includegraphics[width=1.05\hsize]{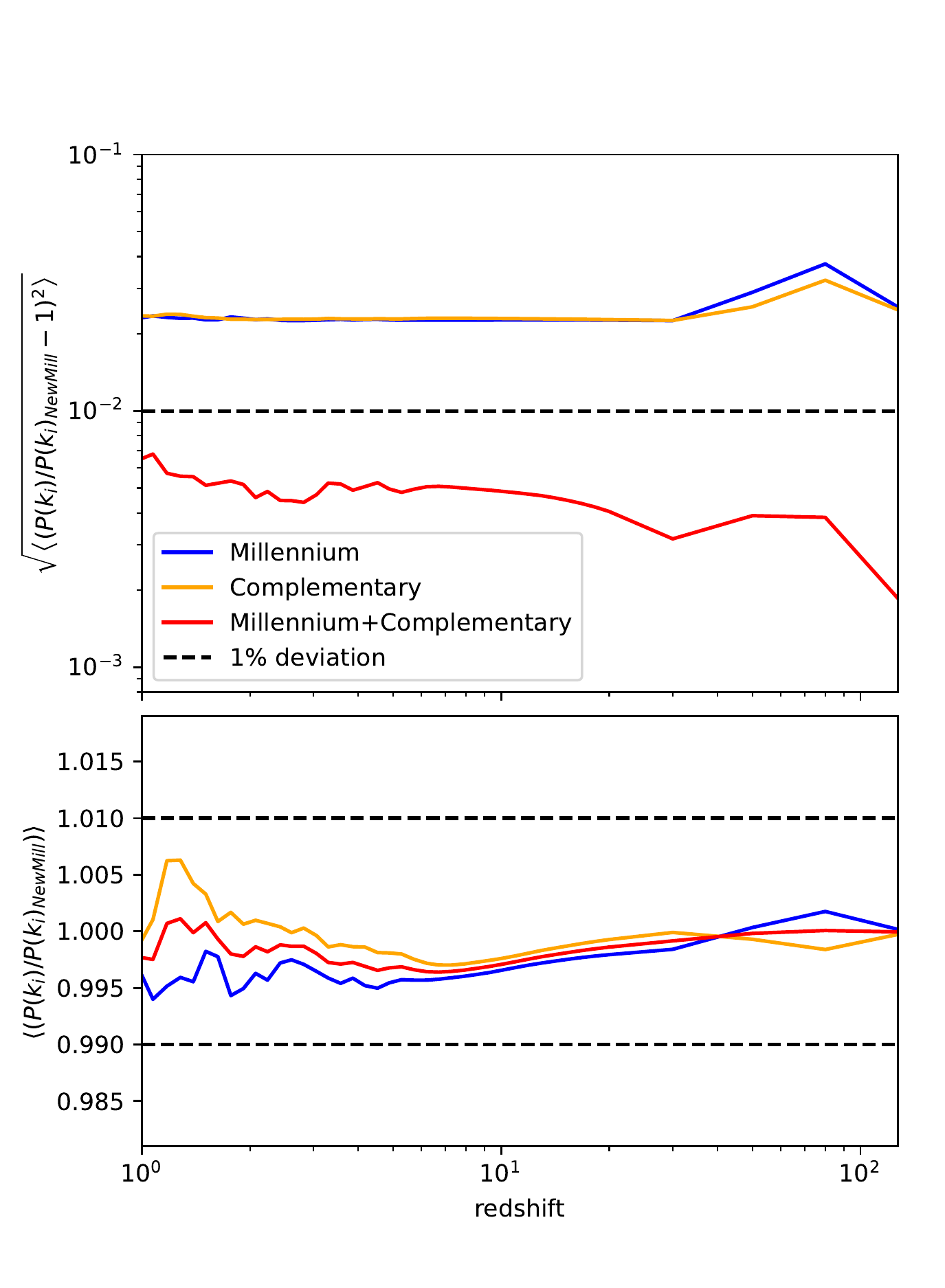}
    \caption{Estimated standard deviation (\textbf{top}) and bias (\textbf{bottom}) of the Millennium simulation and its complementary pair. We used the NewMillennium simulation pair as a reference. For this plot, we used 1100 wave number bins with $\Delta k=9\cdot10^{-3}h/\textnormal{Mpc}$ bin size in $0.0125h/\textnormal{Mpc}<k<10.0h/\textnormal{Mpc}$ range. While individually the two simulations show significant variance in this wave number range and binning, the average of the pair matches the independent NewMillennium average power spectrum with sub-percent accuracy.}
    \label{fig:Mill_vs_NewMill_Pk}
\end{figure}

During the initial condition creation of the complementary pair, we were able to compensate all modes of the original initial condition, because all $k_i$ bins satisfy Eq.~\ref{eq:complementarycriteriaII}. The mass distributions of both simulations are plotted in Fig.~\ref{Fig:DensityField}. The dimensionless power spectrum of both simulation and the estimated cosmic average power spectrum can be seen in Fig.~\ref{Fig:DimlessPk}. To further show that the new simulation method produces consistent results, we compared the power spectrum of the Millennium complementary pair with the average power spectrum of the NewMillennium simulation set in Fig. \ref{fig:Mill_vs_NewMill_Pk}. Using the complementary pair, we estimated the bias of the power spectrum of the original Millennium run due to the initial sampling variance and mode coupling by calculating the weighted mean power spectrum ratio
\begin{equation}
    W = \frac{\sum\limits_{k_{min}<k_i<k_{max}} \Delta N_{m,i} P_{mill}(k_i)/P_{OC}(k_i)}{\sum\limits_{k_{min}<k_i<k_{max}}\Delta N_{m,i}},
\end{equation}
where $P_{mill}(k)$ is the Millennium power spectrum, and $P_{OC}(k)$ is the averaged O-C power spectrum. We find that the original Millennium run, on average, underestimated the power spectrum by $0.997\%$ for $k\leq1.0h/\textnormal{Mpc}$ scales and by $0.0881\%$ for $50h/\textnormal{Mpc}\geq k>1.0h/\textnormal{Mpc}$ wavenumbers at the $z=0$ redshift.

\begin{figure}
    \centering
    \includegraphics[width=1.0\hsize]{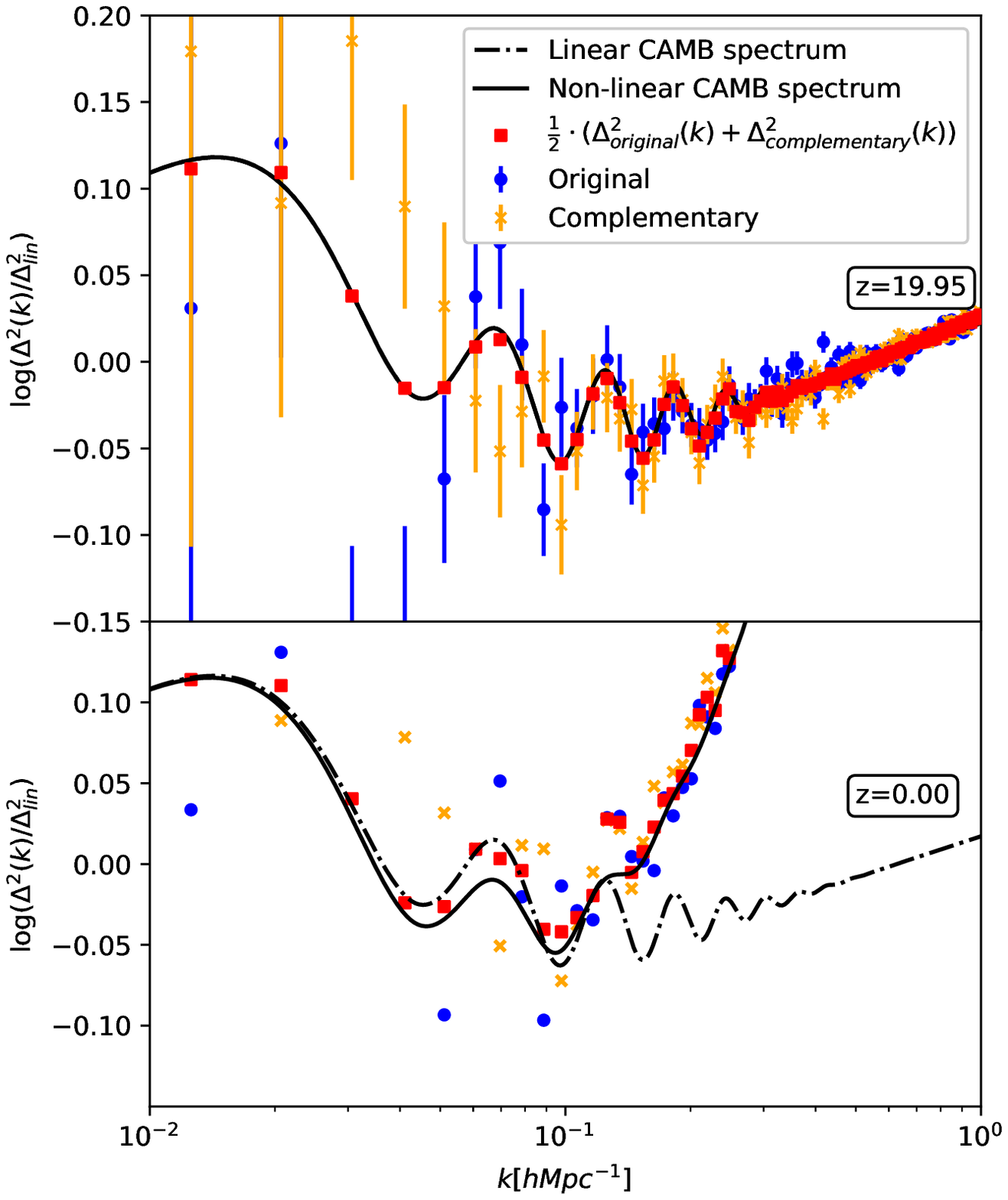}
    \caption{Power spectra of the dark matter density field in the Millennium run and its complementary pair at the BAO scales. All data points have been divided by a linear dark matter-only power spectrum. While the original Millennium simulation cannot resolve these scales by itself due to sample variance, the average spectrum of the original and complementary run can efficiently follow the evolution of the BAO features and matches well at the linear scales with the linear CAMB power spectra.}
    \label{fig:DimlessPkBAO}
\end{figure}

While by itself the original simulation is not large enough to resolve the baryonic wiggles in the power spectrum due to the sample variance at the low-$k$ modes, the average of the two new complementary simulations is able to effectively follow the evolution of these baryon acoustic oscillation (BAO) features, as it can be seen in Fig.~\ref{fig:DimlessPkBAO} with the linear and estimated nonlinear spectrum calculated by Code for Anisotropies in the Microwave Background (CAMB; \cite{2000ApJ...538..473L}). We calculated similar results from the original simulation, but this was only achievable by rescaling the linear initial power spectrum by the estimated scale-dependent nonlinear growth function of the Millennium run.

The new Millennium run and its complementary pair are available to the public on the SciServer platform (URL: \url{https://www.sciserver.org/}) hosted by the Institute for Data Intensive Engineering and Science at the Johns Hopkins University \citep{2020A&C....3300412T}. We made all particle data, halo \citep{1985ApJ...292..371D} and subhalo \citep{2001MNRAS.328..726S} catalogs, and power spectra available here.

\section{Conclusions}
\label{sec:Summary}
   \begin{enumerate}
      \item We reviewed the traditional initial condition generation method, and propose two new techniques for pairs of simulations with reduced cosmic variance: the paired-and-mean and the complementary simulation method.
      \item We compared the power spectrum calculated from the new methods with the paired-and-fixed method, and show that the paired-and-mean and original-complementary pairs produce similar results to the paired-and-fixed method. 
      \item A covariance estimation is left to future work, but based on the fact that the initial amplitudes are not fixed in the new methods, we can expect more precise results than those achievable with the paired-and-fixed method.
      \item We show that complementary pairs for existing simulations can be generated, and the sample-variance errors of the original run can be estimated with this pair due to the fact that structures are evolving in the opposite way in the new simulation.
      \item To demonstrate the effectiveness of the original-complementary method, we generated the complementary pair of the Millennium-run. Using this new simulation, we show that the original simulation underestimates the power spectrum at all scales at the sub-percent level. The average power spectrum of these two simulations is able to directly resolve the BAO features of the power spectrum.
   \end{enumerate}
In this paper, we demonstrated that our new methods can effectively reduce the cosmic variance in N-body simulations. These new methods will be useful in making predictions for future surveys, testing cosmological models, and estimating errors of independent cosmological simulations.

\begin{acknowledgements}
GR’s research was supported by an appointment to the NASA Postdoctoral Program administered by Oak Ridge Associated Universities under contract with NASA. GR and AK were supported by JPL, which is run under contract by California Institute of Technology for NASA.
This work was supported by the  Ministry of Innovation and Technology NRDI Office grants OTKA NN 129148 and the MILAB Artificial Intelligence National Laboratory Program. IS acknowledges support from the National Science Foundation (NSF) award 1616974.
The authors thank Volker Springel for significant help with the original and complementary Millennium initial conditions, and for assistance in the comparison of the different Gadget versions.
We thank Gerard Lemson for making it possible to store the simulation data in the SciServer platform.
The authors acknowledge the Texas Advanced Computing Center (TACC) at The University of Texas at Austin for providing HPC and visualization resources that have contributed to the research results reported within this paper. URL: \url{http://www.tacc.utexas.edu}

\end{acknowledgements}

%
%

\bibliographystyle{aa}
\bibliography{ComplementaryCosmologicalSimulations}

\begin{thebibliography}{50}
\expandafter\ifx\csname natexlab\endcsname\relax\def\natexlab#1{#1}\fi

\bibitem[{{Allgood} {et~al.}(2001){Allgood}, {Blumenthal}, \&
  {Primack}}]{2001astro.ph..9403A}
{Allgood}, B., {Blumenthal}, G., \& {Primack}, J.~R. 2001, arXiv e-prints,
  astro

\bibitem[{{Angulo} \& {Pontzen}(2016)}]{2016MNRAS.462L...1A}
{Angulo}, R.~E. \& {Pontzen}, A. 2016, \mnras, 462, L1

\bibitem[{{Angulo} {et~al.}(2012){Angulo}, {Springel}, {White}, {Jenkins},
  {Baugh}, \& {Frenk}}]{2012MNRAS.426.2046A}
{Angulo}, R.~E., {Springel}, V., {White}, S.~D.~M., {et~al.} 2012, \mnras, 426,
  2046

\bibitem[{{Baugh} \& {Efstathiou}(1994)}]{1994MNRAS.267..323B}
{Baugh}, C.~M. \& {Efstathiou}, G. 1994, \mnras, 267, 323

\bibitem[{{Boylan-Kolchin} {et~al.}(2009){Boylan-Kolchin}, {Springel}, {White},
  {Jenkins}, \& {Lemson}}]{2009MNRAS.398.1150B}
{Boylan-Kolchin}, M., {Springel}, V., {White}, S. D.~M., {Jenkins}, A., \&
  {Lemson}, G. 2009, \mnras, 398, 1150

\bibitem[{{Casarini} {et~al.}(2012){Casarini}, {Bonometto}, {Borgani}, {Dolag},
  {Murante}, {Mezzetti}, {Tornatore}, \& {La Vacca}}]{2012A&A...542A.126C}
{Casarini}, L., {Bonometto}, S.~A., {Borgani}, S., {et~al.} 2012, \aap, 542,
  A126

\bibitem[{{Chambers} {et~al.}(2016){Chambers}, {Magnier}, {Metcalfe},
  {Flewelling}, {Huber}, {Waters}, {Denneau}, {Draper}, {Farrow}, {Finkbeiner},
  {Holmberg}, {Koppenhoefer}, {Price}, {Rest}, {Saglia}, {Schlafly}, {Smartt},
  {Sweeney}, {Wainscoat}, {Burgett}, {Chastel}, {Grav}, {Heasley}, {Hodapp},
  {Jedicke}, {Kaiser}, {Kudritzki}, {Luppino}, {Lupton}, {Monet}, {Morgan},
  {Onaka}, {Shiao}, {Stubbs}, {Tonry}, {White}, {Ba{\~n}ados}, {Bell},
  {Bender}, {Bernard}, {Boegner}, {Boffi}, {Botticella}, {Calamida},
  {Casertano}, {Chen}, {Chen}, {Cole}, {Deacon}, {Frenk}, {Fitzsimmons},
  {Gezari}, {Gibbs}, {Goessl}, {Goggia}, {Gourgue}, {Goldman}, {Grant},
  {Grebel}, {Hambly}, {Hasinger}, {Heavens}, {Heckman}, {Henderson}, {Henning},
  {Holman}, {Hopp}, {Ip}, {Isani}, {Jackson}, {Keyes}, {Koekemoer}, {Kotak},
  {Le}, {Liska}, {Long}, {Lucey}, {Liu}, {Martin}, {Masci}, {McLean}, {Mindel},
  {Misra}, {Morganson}, {Murphy}, {Obaika}, {Narayan}, {Nieto-Santisteban},
  {Norberg}, {Peacock}, {Pier}, {Postman}, {Primak}, {Rae}, {Rai}, {Riess},
  {Riffeser}, {Rix}, {R{\"o}ser}, {Russel}, {Rutz}, {Schilbach}, {Schultz},
  {Scolnic}, {Strolger}, {Szalay}, {Seitz}, {Small}, {Smith}, {Soderblom},
  {Taylor}, {Thomson}, {Taylor}, {Thakar}, {Thiel}, {Thilker}, {Unger},
  {Urata}, {Valenti}, {Wagner}, {Walder}, {Walter}, {Watters}, {Werner},
  {Wood-Vasey}, \& {Wyse}}]{2016arXiv161205560C}
{Chambers}, K.~C., {Magnier}, E.~A., {Metcalfe}, N., {et~al.} 2016, arXiv
  e-prints, arXiv:1612.05560

\bibitem[{{Crocce} {et~al.}(2006){Crocce}, {Pueblas}, \&
  {Scoccimarro}}]{2006MNRAS.373..369C}
{Crocce}, M., {Pueblas}, S., \& {Scoccimarro}, R. 2006, \mnras, 373, 369

\bibitem[{{Crocce} {et~al.}(2012){Crocce}, {Pueblas}, \&
  {Scoccimarro}}]{2012ascl.soft01005C}
{Crocce}, M., {Pueblas}, S., \& {Scoccimarro}, R. 2012, {2LPTIC: 2nd-order
  Lagrangian Perturbation Theory Initial Conditions}

\bibitem[{{Davis} {et~al.}(1985){Davis}, {Efstathiou}, {Frenk}, \&
  {White}}]{1985ApJ...292..371D}
{Davis}, M., {Efstathiou}, G., {Frenk}, C.~S., \& {White}, S.~D.~M. 1985, \apj,
  292, 371

\bibitem[{{DESI Collaboration} {et~al.}(2016){DESI Collaboration}, {Aghamousa},
  {Aguilar}, {Ahlen}, {Alam}, {Allen}, {Allende Prieto}, {Annis}, {Bailey},
  {Balland}, {Ballester}, {Baltay}, {Beaufore}, {Bebek}, {Beers}, {Bell},
  {Bernal}, {Besuner}, {Beutler}, {Blake}, {Bleuler}, {Blomqvist}, {Blum},
  {Bolton}, {Briceno}, {Brooks}, {Brownstein}, {Buckley-Geer}, {Burden},
  {Burtin}, {Busca}, {Cahn}, {Cai}, {Cardiel-Sas}, {Carlberg}, {Carton},
  {Casas}, {Castander}, {Cervantes-Cota}, {Claybaugh}, {Close}, {Coker},
  {Cole}, {Comparat}, {Cooper}, {Cousinou}, {Crocce}, {Cuby}, {Cunningham},
  {Davis}, {Dawson}, {de la Macorra}, {De Vicente}, {Delubac}, {Derwent},
  {Dey}, {Dhungana}, {Ding}, {Doel}, {Duan}, {Ealet}, {Edelstein},
  {Eftekharzadeh}, {Eisenstein}, {Elliott}, {Escoffier}, {Evatt}, {Fagrelius},
  {Fan}, {Fanning}, {Farahi}, {Farihi}, {Favole}, {Feng}, {Fernandez},
  {Findlay}, {Finkbeiner}, {Fitzpatrick}, {Flaugher}, {Flender}, {Font-Ribera},
  {Forero-Romero}, {Fosalba}, {Frenk}, {Fumagalli}, {Gaensicke}, {Gallo},
  {Garcia-Bellido}, {Gaztanaga}, {Pietro Gentile Fusillo}, {Gerard},
  {Gershkovich}, {Giannantonio}, {Gillet}, {Gonzalez-de-Rivera},
  {Gonzalez-Perez}, {Gott}, {Graur}, {Gutierrez}, {Guy}, {Habib}, {Heetderks},
  {Heetderks}, {Heitmann}, {Hellwing}, {Herrera}, {Ho}, {Holland}, {Honscheid},
  {Huff}, {Hutchinson}, {Huterer}, {Hwang}, {Illa Laguna}, {Ishikawa},
  {Jacobs}, {Jeffrey}, {Jelinsky}, {Jennings}, {Jiang}, {Jimenez}, {Johnson},
  {Joyce}, {Jullo}, {Juneau}, {Kama}, {Karcher}, {Karkar}, {Kehoe}, {Kennamer},
  {Kent}, {Kilbinger}, {Kim}, {Kirkby}, {Kisner}, {Kitanidis}, {Kneib},
  {Koposov}, {Kovacs}, {Koyama}, {Kremin}, {Kron}, {Kronig}, {Kueter-Young},
  {Lacey}, {Lafever}, {Lahav}, {Lambert}, {Lampton}, {Landriau}, {Lang},
  {Lauer}, {Le Goff}, {Le Guillou}, {Le Van Suu}, {Lee}, {Lee}, {Leitner},
  {Lesser}, {Levi}, {L'Huillier}, {Li}, {Liang}, {Lin}, {Linder}, {Loebman},
  {Luki{\'c}}, {Ma}, {MacCrann}, {Magneville}, {Makarem}, {Manera}, {Manser},
  {Marshall}, {Martini}, {Massey}, {Matheson}, {McCauley}, {McDonald},
  {McGreer}, {Meisner}, {Metcalfe}, {Miller}, {Miquel}, {Moustakas}, {Myers},
  {Naik}, {Newman}, {Nichol}, {Nicola}, {Nicolati da Costa}, {Nie}, {Niz},
  {Norberg}, {Nord}, {Norman}, {Nugent}, {O'Brien}, {Oh}, {Olsen}, {Padilla},
  {Padmanabhan}, {Padmanabhan}, {Palanque-Delabrouille}, {Palmese},
  {Pappalardo}, {P{\^a}ris}, {Park}, {Patej}, {Peacock}, {Peiris}, {Peng},
  {Percival}, {Perruchot}, {Pieri}, {Pogge}, {Pollack}, {Poppett}, {Prada},
  {Prakash}, {Probst}, {Rabinowitz}, {Raichoor}, {Ree}, {Refregier}, {Regal},
  {Reid}, {Reil}, {Rezaie}, {Rockosi}, {Roe}, {Ronayette}, {Roodman}, {Ross},
  {Ross}, {Rossi}, {Rozo}, {Ruhlmann-Kleider}, {Rykoff}, {Sabiu}, {Samushia},
  {Sanchez}, {Sanchez}, {Schlegel}, {Schneider}, {Schubnell}, {Secroun},
  {Seljak}, {Seo}, {Serrano}, {Shafieloo}, {Shan}, {Sharples}, {Sholl},
  {Shourt}, {Silber}, {Silva}, {Sirk}, {Slosar}, {Smith}, {Smoot}, {Som},
  {Song}, {Sprayberry}, {Staten}, {Stefanik}, {Tarle}, {Sien Tie}, {Tinker},
  {Tojeiro}, {Valdes}, {Valenzuela}, {Valluri}, {Vargas-Magana}, {Verde},
  {Walker}, {Wang}, {Wang}, {Weaver}, {Weaverdyck}, {Wechsler}, {Weinberg},
  {White}, {Yang}, {Yeche}, {Zhang}, {Zhao}, {Zheng}, {Zhou}, {Zhou}, {Zhu},
  {Zou}, \& {Zu}}]{2016arXiv161100036D}
{DESI Collaboration}, {Aghamousa}, A., {Aguilar}, J., {et~al.} 2016, arXiv
  e-prints, arXiv:1611.00036

\bibitem[{{Dor{\'e}} {et~al.}(2014){Dor{\'e}}, {Bock}, {Ashby}, {Capak},
  {Cooray}, {de Putter}, {Eifler}, {Flagey}, {Gong}, {Habib}, {Heitmann},
  {Hirata}, {Jeong}, {Katti}, {Korngut}, {Krause}, {Lee}, {Masters},
  {Mauskopf}, {Melnick}, {Mennesson}, {Nguyen}, {{\"O}berg}, {Pullen},
  {Raccanelli}, {Smith}, {Song}, {Tolls}, {Unwin}, {Venumadhav}, {Viero},
  {Werner}, \& {Zemcov}}]{2014arXiv1412.4872D}
{Dor{\'e}}, O., {Bock}, J., {Ashby}, M., {et~al.} 2014, arXiv e-prints,
  arXiv:1412.4872

\bibitem[{{Driver} \& {Robotham}(2010)}]{2010MNRAS.407.2131D}
{Driver}, S.~P. \& {Robotham}, A. S.~G. 2010, \mnras, 407, 2131

\bibitem[{{Falck} {et~al.}(2021){Falck}, {Wang}, {Jenkins}, {Lemson},
  {Medvedev}, {Neyrinck}, \& {Szalay}}]{2021MNRAS.506.2659F}
{Falck}, B., {Wang}, J., {Jenkins}, A., {et~al.} 2021, \mnras, 506, 2659

\bibitem[{{Green} {et~al.}(2012){Green}, {Schechter}, {Baltay}, {Bean},
  {Bennett}, {Brown}, {Conselice}, {Donahue}, {Fan}, {Gaudi}, {Hirata},
  {Kalirai}, {Lauer}, {Nichol}, {Padmanabhan}, {Perlmutter}, {Rauscher},
  {Rhodes}, {Roellig}, {Stern}, {Sumi}, {Tanner}, {Wang}, {Weinberg}, {Wright},
  {Gehrels}, {Sambruna}, {Traub}, {Anderson}, {Cook}, {Garnavich},
  {Hillenbrand}, {Ivezic}, {Kerins}, {Lunine}, {McDonald}, {Penny}, {Phillips},
  {Rieke}, {Riess}, {van der Marel}, {Barry}, {Cheng}, {Content}, {Cutri},
  {Goullioud}, {Grady}, {Helou}, {Jackson}, {Kruk}, {Melton}, {Peddie},
  {Rioux}, \& {Seiffert}}]{2012arXiv1208.4012G}
{Green}, J., {Schechter}, P., {Baltay}, C., {et~al.} 2012, arXiv e-prints,
  arXiv:1208.4012

\bibitem[{{Hamilton} {et~al.}(2006){Hamilton}, {Rimes}, \&
  {Scoccimarro}}]{2006MNRAS.371.1188H}
{Hamilton}, A. J.~S., {Rimes}, C.~D., \& {Scoccimarro}, R. 2006, \mnras, 371,
  1188

\bibitem[{{Harnois-D{\'e}raps} {et~al.}(2019){Harnois-D{\'e}raps}, {Giblin}, \&
  {Joachimi}}]{2019A&A...631A.160H}
{Harnois-D{\'e}raps}, J., {Giblin}, B., \& {Joachimi}, B. 2019, \aap, 631, A160

\bibitem[{{Heitmann} {et~al.}(2019){Heitmann}, {Finkel}, {Pope}, {Morozov},
  {Frontiere}, {Habib}, {Rangel}, {Uram}, {Korytov}, {Child}, {Flender},
  {Insley}, \& {Rizzi}}]{2019ApJS..245...16H}
{Heitmann}, K., {Finkel}, H., {Pope}, A., {et~al.} 2019, \apjs, 245, 16

\bibitem[{{Hockney} \& {Eastwood}(1988)}]{1988csup.book.....H}
{Hockney}, R.~W. \& {Eastwood}, J.~W. 1988, {Computer simulation using
  particles}

\bibitem[{{Jenkins}(2010)}]{2010MNRAS.403.1859J}
{Jenkins}, A. 2010, \mnras, 403, 1859

\bibitem[{{Jenkins} {et~al.}(2001){Jenkins}, {Frenk}, {White}, {Colberg},
  {Cole}, {Evrard}, {Couchman}, \& {Yoshida}}]{2001MNRAS.321..372J}
{Jenkins}, A., {Frenk}, C.~S., {White}, S.~D.~M., {et~al.} 2001, \mnras, 321,
  372

\bibitem[{{Klypin} \& {Holtzman}(1997)}]{1997astro.ph.12217K}
{Klypin}, A. \& {Holtzman}, J. 1997, arXiv e-prints, astro

\bibitem[{{Klypin} {et~al.}(2020){Klypin}, {Prada}, \&
  {Byun}}]{2020MNRAS.496.3862K}
{Klypin}, A., {Prada}, F., \& {Byun}, J. 2020, \mnras, 496, 3862

\bibitem[{{Klypin} {et~al.}(2016){Klypin}, {Yepes}, {Gottl{\"o}ber}, {Prada},
  \& {He{\ss}}}]{2016MNRAS.457.4340K}
{Klypin}, A., {Yepes}, G., {Gottl{\"o}ber}, S., {Prada}, F., \& {He{\ss}}, S.
  2016, \mnras, 457, 4340

\bibitem[{{Klypin} {et~al.}(2011){Klypin}, {Trujillo-Gomez}, \&
  {Primack}}]{2011ApJ...740..102K}
{Klypin}, A.~A., {Trujillo-Gomez}, S., \& {Primack}, J. 2011, \apj, 740, 102

\bibitem[{{Lewis} {et~al.}(2000){Lewis}, {Challinor}, \&
  {Lasenby}}]{2000ApJ...538..473L}
{Lewis}, A., {Challinor}, A., \& {Lasenby}, A. 2000, \apj, 538, 473

\bibitem[{{LSST Science Collaboration} {et~al.}(2009){LSST Science
  Collaboration}, {Abell}, {Allison}, {Anderson}, {Andrew}, {Angel}, {Armus},
  {Arnett}, {Asztalos}, {Axelrod}, {Bailey}, {Ballantyne}, {Bankert},
  {Barkhouse}, {Barr}, {Barrientos}, {Barth}, {Bartlett}, {Becker}, {Becla},
  {Beers}, {Bernstein}, {Biswas}, {Blanton}, {Bloom}, {Bochanski}, {Boeshaar},
  {Borne}, {Bradac}, {Brandt}, {Bridge}, {Brown}, {Brunner}, {Bullock},
  {Burgasser}, {Burge}, {Burke}, {Cargile}, {Chandrasekharan}, {Chartas},
  {Chesley}, {Chu}, {Cinabro}, {Claire}, {Claver}, {Clowe}, {Connolly}, {Cook},
  {Cooke}, {Cooray}, {Covey}, {Culliton}, {de Jong}, {de Vries}, {Debattista},
  {Delgado}, {Dell'Antonio}, {Dhital}, {Di Stefano}, {Dickinson}, {Dilday},
  {Djorgovski}, {Dobler}, {Donalek}, {Dubois-Felsmann}, {Durech},
  {Eliasdottir}, {Eracleous}, {Eyer}, {Falco}, {Fan}, {Fassnacht}, {Ferguson},
  {Fernandez}, {Fields}, {Finkbeiner}, {Figueroa}, {Fox}, {Francke}, {Frank},
  {Frieman}, {Fromenteau}, {Furqan}, {Galaz}, {Gal-Yam}, {Garnavich},
  {Gawiser}, {Geary}, {Gee}, {Gibson}, {Gilmore}, {Grace}, {Green}, {Gressler},
  {Grillmair}, {Habib}, {Haggerty}, {Hamuy}, {Harris}, {Hawley}, {Heavens},
  {Hebb}, {Henry}, {Hileman}, {Hilton}, {Hoadley}, {Holberg}, {Holman},
  {Howell}, {Infante}, {Ivezic}, {Jacoby}, {Jain}, {R}, {Jedicke}, {Jee},
  {Garrett Jernigan}, {Jha}, {Johnston}, {Jones}, {Juric}, {Kaasalainen},
  {Styliani}, {Kafka}, {Kahn}, {Kaib}, {Kalirai}, {Kantor}, {Kasliwal},
  {Keeton}, {Kessler}, {Knezevic}, {Kowalski}, {Krabbendam}, {Krughoff},
  {Kulkarni}, {Kuhlman}, {Lacy}, {Lepine}, {Liang}, {Lien}, {Lira}, {Long},
  {Lorenz}, {Lotz}, {Lupton}, {Lutz}, {Macri}, {Mahabal}, {Mandelbaum},
  {Marshall}, {May}, {McGehee}, {Meadows}, {Meert}, {Milani}, {Miller},
  {Miller}, {Mills}, {Minniti}, {Monet}, {Mukadam}, {Nakar}, {Neill}, {Newman},
  {Nikolaev}, {Nordby}, {O'Connor}, {Oguri}, {Oliver}, {Olivier}, {Olsen},
  {Olsen}, {Olszewski}, {Oluseyi}, {Padilla}, {Parker}, {Pepper}, {Peterson},
  {Petry}, {Pinto}, {Pizagno}, {Popescu}, {Prsa}, {Radcka}, {Raddick},
  {Rasmussen}, {Rau}, {Rho}, {Rhoads}, {Richards}, {Ridgway}, {Robertson},
  {Roskar}, {Saha}, {Sarajedini}, {Scannapieco}, {Schalk}, {Schindler},
  {Schmidt}, {Schmidt}, {Schneider}, {Schumacher}, {Scranton}, {Sebag},
  {Seppala}, {Shemmer}, {Simon}, {Sivertz}, {Smith}, {Allyn Smith}, {Smith},
  {Spitz}, {Stanford}, {Stassun}, {Strader}, {Strauss}, {Stubbs}, {Sweeney},
  {Szalay}, {Szkody}, {Takada}, {Thorman}, {Trilling}, {Trimble}, {Tyson}, {Van
  Berg}, {Vanden Berk}, {VanderPlas}, {Verde}, {Vrsnak}, {Walkowicz},
  {Wandelt}, {Wang}, {Wang}, {Warner}, {Wechsler}, {West}, {Wiecha},
  {Williams}, {Willman}, {Wittman}, {Wolff}, {Wood-Vasey}, {Wozniak}, {Young},
  {Zentner}, \& {Zhan}}]{2009arXiv0912.0201L}
{LSST Science Collaboration}, {Abell}, P.~A., {Allison}, J., {et~al.} 2009,
  arXiv e-prints, arXiv:0912.0201

\bibitem[{{Maddox} {et~al.}(1990){Maddox}, {Sutherland}, {Efstathiou}, \&
  {Loveday}}]{1990MNRAS.243..692M}
{Maddox}, S.~J., {Sutherland}, W.~J., {Efstathiou}, G., \& {Loveday}, J. 1990,
  \mnras, 243, 692

\bibitem[{{Maksimova} {et~al.}(2021){Maksimova}, {Garrison}, {Eisenstein},
  {Hadzhiyska}, {Bose}, \& {Satterthwaite}}]{2021MNRAS.508.4017M}
{Maksimova}, N.~A., {Garrison}, L.~H., {Eisenstein}, D.~J., {et~al.} 2021,
  \mnras, 508, 4017

\bibitem[{{McBride} {et~al.}(2009){McBride}, {Berlind}, {Scoccimarro},
  {Wechsler}, {Busha}, {Gardner}, \& {van den Bosch}}]{2009AAS...21342506M}
{McBride}, C., {Berlind}, A., {Scoccimarro}, R., {et~al.} 2009, in American
  Astronomical Society Meeting Abstracts, Vol. 213, American Astronomical
  Society Meeting Abstracts \#213, 425.06

\bibitem[{{Moster} {et~al.}(2011){Moster}, {Somerville}, {Newman}, \&
  {Rix}}]{2011ApJ...731..113M}
{Moster}, B.~P., {Somerville}, R.~S., {Newman}, J.~A., \& {Rix}, H.-W. 2011,
  \apj, 731, 113

\bibitem[{{Percival} {et~al.}(2001){Percival}, {Baugh}, {Bland-Hawthorn},
  {Bridges}, {Cannon}, {Cole}, {Colless}, {Collins}, {Couch}, {Dalton}, {De
  Propris}, {Driver}, {Efstathiou}, {Ellis}, {Frenk}, {Glazebrook}, {Jackson},
  {Lahav}, {Lewis}, {Lumsden}, {Maddox}, {Moody}, {Norberg}, {Peacock},
  {Peterson}, {Sutherland}, \& {Taylor}}]{2001MNRAS.327.1297P}
{Percival}, W.~J., {Baugh}, C.~M., {Bland-Hawthorn}, J., {et~al.} 2001, \mnras,
  327, 1297

\bibitem[{{Planck Collaboration} {et~al.}(2020){Planck Collaboration},
  {Aghanim}, {Akrami}, {Ashdown}, {Aumont}, {Baccigalupi}, {Ballardini},
  {Banday}, {Barreiro}, {Bartolo}, {Basak}, {Battye}, {Benabed}, {Bernard},
  {Bersanelli}, {Bielewicz}, {Bock}, {Bond}, {Borrill}, {Bouchet}, {Boulanger},
  {Bucher}, {Burigana}, {Butler}, {Calabrese}, {Cardoso}, {Carron},
  {Challinor}, {Chiang}, {Chluba}, {Colombo}, {Combet}, {Contreras}, {Crill},
  {Cuttaia}, {de Bernardis}, {de Zotti}, {Delabrouille}, {Delouis}, {Di
  Valentino}, {Diego}, {Dor{\'e}}, {Douspis}, {Ducout}, {Dupac}, {Dusini},
  {Efstathiou}, {Elsner}, {En{\ss}lin}, {Eriksen}, {Fantaye}, {Farhang},
  {Fergusson}, {Fernandez-Cobos}, {Finelli}, {Forastieri}, {Frailis},
  {Fraisse}, {Franceschi}, {Frolov}, {Galeotta}, {Galli}, {Ganga},
  {G{\'e}nova-Santos}, {Gerbino}, {Ghosh}, {Gonz{\'a}lez-Nuevo}, {G{\'o}rski},
  {Gratton}, {Gruppuso}, {Gudmundsson}, {Hamann}, {Handley}, {Hansen},
  {Herranz}, {Hildebrandt}, {Hivon}, {Huang}, {Jaffe}, {Jones}, {Karakci},
  {Keih{\"a}nen}, {Keskitalo}, {Kiiveri}, {Kim}, {Kisner}, {Knox},
  {Krachmalnicoff}, {Kunz}, {Kurki-Suonio}, {Lagache}, {Lamarre}, {Lasenby},
  {Lattanzi}, {Lawrence}, {Le Jeune}, {Lemos}, {Lesgourgues}, {Levrier},
  {Lewis}, {Liguori}, {Lilje}, {Lilley}, {Lindholm}, {L{\'o}pez-Caniego},
  {Lubin}, {Ma}, {Mac{\'\i}as-P{\'e}rez}, {Maggio}, {Maino}, {Mandolesi},
  {Mangilli}, {Marcos-Caballero}, {Maris}, {Martin}, {Martinelli},
  {Mart{\'\i}nez-Gonz{\'a}lez}, {Matarrese}, {Mauri}, {McEwen}, {Meinhold},
  {Melchiorri}, {Mennella}, {Migliaccio}, {Millea}, {Mitra},
  {Miville-Desch{\^e}nes}, {Molinari}, {Montier}, {Morgante}, {Moss}, {Natoli},
  {N{\o}rgaard-Nielsen}, {Pagano}, {Paoletti}, {Partridge}, {Patanchon},
  {Peiris}, {Perrotta}, {Pettorino}, {Piacentini}, {Polastri}, {Polenta},
  {Puget}, {Rachen}, {Reinecke}, {Remazeilles}, {Renzi}, {Rocha}, {Rosset},
  {Roudier}, {Rubi{\~n}o-Mart{\'\i}n}, {Ruiz-Granados}, {Salvati}, {Sandri},
  {Savelainen}, {Scott}, {Shellard}, {Sirignano}, {Sirri}, {Spencer},
  {Sunyaev}, {Suur-Uski}, {Tauber}, {Tavagnacco}, {Tenti}, {Toffolatti},
  {Tomasi}, {Trombetti}, {Valenziano}, {Valiviita}, {Van Tent}, {Vibert},
  {Vielva}, {Villa}, {Vittorio}, {Wandelt}, {Wehus}, {White}, {White},
  {Zacchei}, \& {Zonca}}]{2020A&A...641A...6P}
{Planck Collaboration}, {Aghanim}, N., {Akrami}, Y., {et~al.} 2020, \aap, 641,
  A6

\bibitem[{{Pontzen} {et~al.}(2016){Pontzen}, {Slosar}, {Roth}, \&
  {Peiris}}]{2016PhRvD..93j3519P}
{Pontzen}, A., {Slosar}, A., {Roth}, N., \& {Peiris}, H.~V. 2016, \prd, 93,
  103519

\bibitem[{{Potter} {et~al.}(2017){Potter}, {Stadel}, \&
  {Teyssier}}]{2017ComAC...4....2P}
{Potter}, D., {Stadel}, J., \& {Teyssier}, R. 2017, Computational Astrophysics
  and Cosmology, 4, 2

\bibitem[{{Rimes} \& {Hamilton}(2006)}]{2006MNRAS.371.1205R}
{Rimes}, C.~D. \& {Hamilton}, A. J.~S. 2006, \mnras, 371, 1205

\bibitem[{{Schneider} {et~al.}(2016){Schneider}, {Teyssier}, {Potter},
  {Stadel}, {Onions}, {Reed}, {Smith}, {Springel}, {Pearce}, \&
  {Scoccimarro}}]{2016JCAP...04..047S}
{Schneider}, A., {Teyssier}, R., {Potter}, D., {et~al.} 2016, \jcap, 2016, 047

\bibitem[{{Sirko}(2005)}]{2005ApJ...634..728S}
{Sirko}, E. 2005, \apj, 634, 728

\bibitem[{{Springel}(2005)}]{2005MNRAS.364.1105S}
{Springel}, V. 2005, \mnras, 364, 1105

\bibitem[{{Springel}(2015)}]{2015ascl.soft02003S}
{Springel}, V. 2015, {N-GenIC: Cosmological structure initial conditions},
  Astrophysics Source Code Library, record ascl:1502.003

\bibitem[{{Springel} {et~al.}(2021){Springel}, {Pakmor}, {Zier}, \&
  {Reinecke}}]{2021MNRAS.506.2871S}
{Springel}, V., {Pakmor}, R., {Zier}, O., \& {Reinecke}, M. 2021, \mnras, 506,
  2871

\bibitem[{{Springel} {et~al.}(2005){Springel}, {White}, {Jenkins}, {Frenk},
  {Yoshida}, {Gao}, {Navarro}, {Thacker}, {Croton}, {Helly}, {Peacock}, {Cole},
  {Thomas}, {Couchman}, {Evrard}, {Colberg}, \& {Pearce}}]{2005Natur.435..629S}
{Springel}, V., {White}, S. D.~M., {Jenkins}, A., {et~al.} 2005, \nat, 435, 629

\bibitem[{{Springel} {et~al.}(2001){Springel}, {White}, {Tormen}, \&
  {Kauffmann}}]{2001MNRAS.328..726S}
{Springel}, V., {White}, S. D.~M., {Tormen}, G., \& {Kauffmann}, G. 2001,
  \mnras, 328, 726

\bibitem[{{Taghizadeh-Popp} {et~al.}(2020){Taghizadeh-Popp}, {Kim}, {Lemson},
  {Medvedev}, {Raddick}, {Szalay}, {Thakar}, {Booker}, {Chhetri}, {Dobos}, \&
  {Rippin}}]{2020A&C....3300412T}
{Taghizadeh-Popp}, M., {Kim}, J.~W., {Lemson}, G., {et~al.} 2020, Astronomy and
  Computing, 33, 100412

\bibitem[{{Tegmark} {et~al.}(2004){Tegmark}, {Blanton}, {Strauss}, {Hoyle},
  {Schlegel}, {Scoccimarro}, {Vogeley}, {Weinberg}, {Zehavi}, {Berlind},
  {Budavari}, {Connolly}, {Eisenstein}, {Finkbeiner}, {Frieman}, {Gunn},
  {Hamilton}, {Hui}, {Jain}, {Johnston}, {Kent}, {Lin}, {Nakajima}, {Nichol},
  {Ostriker}, {Pope}, {Scranton}, {Seljak}, {Sheth}, {Stebbins}, {Szalay},
  {Szapudi}, {Verde}, {Xu}, {Annis}, {Bahcall}, {Brinkmann}, {Burles},
  {Castander}, {Csabai}, {Loveday}, {Doi}, {Fukugita}, {Gott}, {Hennessy},
  {Hogg}, {Ivezi{\'c}}, {Knapp}, {Lamb}, {Lee}, {Lupton}, {McKay}, {Kunszt},
  {Munn}, {O'Connell}, {Peoples}, {Pier}, {Richmond}, {Rockosi}, {Schneider},
  {Stoughton}, {Tucker}, {Vanden Berk}, {Yanny}, {York}, \& {SDSS
  Collaboration}}]{2004ApJ...606..702T}
{Tegmark}, M., {Blanton}, M.~R., {Strauss}, M.~A., {et~al.} 2004, \apj, 606,
  702

\bibitem[{{The Dark Energy Survey Collaboration}(2005)}]{2005astro.ph.10346T}
{The Dark Energy Survey Collaboration}. 2005, arXiv e-prints, astro

\bibitem[{{Tutusaus} {et~al.}(2020){Tutusaus}, {Martinelli}, {Cardone},
  {Camera}, {Yahia-Cherif}, {Casas}, {Blanchard}, {Kilbinger}, {Lacasa},
  {Sakr}, {Ili{\'c}}, {Kunz}, {Carbone}, {Castander}, {Dournac}, {Fosalba},
  {Kitching}, {Markovic}, {Mangilli}, {Pettorino}, {Sapone}, {Yankelevich},
  {Auricchio}, {Bender}, {Bonino}, {Boucaud}, {Brescia}, {Capobianco},
  {Carretero}, {Castellano}, {Cavuoti}, {Cledassou}, {Congedo}, {Conversi},
  {Corcione}, {Costille}, {Crocce}, {Cropper}, {Dubath}, {Dusini}, {Fabbian},
  {Frailis}, {Franceschi}, {Garilli}, {Grupp}, {Guzzo}, {Hoekstra}, {Hormuth},
  {Israel}, {Jahnke}, {Kermiche}, {Kubik}, {Laureijs}, {Ligori}, {Lilje},
  {Lloro}, {Maiorano}, {Marggraf}, {Massey}, {Mei}, {Merlin}, {Meylan},
  {Moscardini}, {Ntelis}, {Padilla}, {Paltani}, {Pasian}, {Percival}, {Pires},
  {Poncet}, {Raison}, {Rhodes}, {Roncarelli}, {Rossetti}, {Saglia},
  {Schneider}, {Secroun}, {Serrano}, {Sirignano}, {Sirri}, {Starck}, {Sureau},
  {Taylor}, {Tereno}, {Toledo-Moreo}, {Valenziano}, {Wang}, {Welikala},
  {Weller}, {Zacchei}, \& {Zoubian}}]{2020A&A...643A..70T}
{Tutusaus}, I., {Martinelli}, M., {Cardone}, V.~F., {et~al.} 2020, \aap, 643,
  A70

\bibitem[{{Valdarnini} \& {Borgani}(1991)}]{1991MNRAS.251..575V}
{Valdarnini}, R. \& {Borgani}, S. 1991, \mnras, 251, 575

\bibitem[{{Villaescusa-Navarro} {et~al.}(2018){Villaescusa-Navarro}, {Naess},
  {Genel}, {Pontzen}, {Wandelt}, {Anderson}, {Font-Ribera}, {Battaglia}, \&
  {Spergel}}]{2018ApJ...867..137V}
{Villaescusa-Navarro}, F., {Naess}, S., {Genel}, S., {et~al.} 2018, \apj, 867,
  137

\bibitem[{{Zel'dovich}(1970)}]{1970A&A.....5...84Z}
{Zel'dovich}, Y.~B. 1970, \aap, 5, 84

\end{thebibliography}

\end{document}